\documentclass[
reprint,
superscriptaddress,
floatfix]{revtex4-2}

\usepackage{amsmath}
\usepackage{amssymb}
\usepackage{graphicx}
\usepackage{dcolumn}
\usepackage{bm}
\usepackage{mathrsfs}
\usepackage{hyperref}
\begin{document}

\title{Negative-energy spin waves in antiferromagnets for spin-current amplification and analogue gravity}   

\author{V.Errani}
\affiliation{
Institute for Theoretical Physics, Utrecht University, 3584CC Utrecht, The Netherlands
}
\author{S.H. Schoenmaker}
\affiliation{Institut f\"ur Physik, Johannes Gutenberg Universit\"at Mainz, 55128, Mainz, Germany}
\author{X.R. Wang}
\affiliation{
Physics Department, The Hong Kong University of Science and Technology (HKUST), Clear Water Bay, Kowloon, Hong Kong}
\affiliation{HKUST Shenzhen Research Institute, Shenzhen 518057, China}
\author{O. Gomonay}
\affiliation{Institut für Physik, Johannes Gutenberg Universität Mainz, 55128, Mainz, Germany}
\author{R.A.Duine}
\affiliation{
Department of Applied Physics, Eindhoven University of Technology,
P.O. Box 513, 5600 MB Eindhoven, The Netherlands}
\affiliation{
Institute for Theoretical Physics, Utrecht University, 3584CC Utrecht, The Netherlands
}
\date{\today}

\begin{abstract}

Magnonic black holes—analogue event horizons for the spin-wave collective excitations of ordered magnets—can be used for
fundamental research, for example for investigating Hawking radiation, but also for technological applications of spin waves. Here we show how to engineer a horizon for spin waves in antiferromagnets, which have the attractive feature of fast magnetization dynamics and linear dispersion relation. We propose a set-up with spatially varying exchange interaction with spin transfer torque to implement the horizon and a second set-up for the amplification of spin waves consisting of an antiferromagnet subject to a spatially varying external magnetic field that is driven by spin orbit torque. We compute the values of parameters needed to implement the horizon and to have amplification
of spin waves. We develop the corresponding Klein-Gordon equation and quantify the amplification. 
Our work paves the way for investigation of Hawking radiation of spin waves and for antiferromagnet-based spin-waves amplifiers.
\end{abstract}
\maketitle
\section{Introduction}
Energy efficiency is one of the problems our society is challenged with, as CMOS technology has already reached its limits in terms of miniaturization and performance~\cite{inproceedings,markov2014limits,haron2008cmos}. One therefore needs to search for new platforms to transport information that are not solely based on the charge of electrons. For this purpose, the spin degree of freedom of electrons is exploited. This is the aim of spintronics. Spin waves are potentially energy-efficient carriers of spin information because of the absence of Joule heating in their propagation. Spin waves are particularly suitable in comparison with, for example, photons, because of the shorter wavelength which is an asset for the purpose of miniaturization.

It is possible to manipulate the dispersion relation of spin waves as it depends on several parameters, such as magnetic fields, anisotropies and exchange interactions. 
Creating a spin wave costs energy, as it is an excitation above the magnetic ground state.
Spin waves which gain energy, i.e., that lower the total energy if they are excited, are also possible. Such negative-energy spin waves would occur if the system is excited from a metastable state rather than the true ground state. Without a stabilizing mechanism, any metastable state would, however, be unstable in the presence of dissipation.
The dispersion relation of spin waves is also modified by spin current, and, in particular, two mechanisms for this are spin transfer torque (STT)~\cite{stiles2002anatomy,duine2009generation,stiles2006spin,gomonay2010spin} and spin orbit torque (SOT)~\cite{manchon2019current,brataas2014spin,demidov2020spin}. Through those two mechanisms it is possible to dynamically stabilize the system, making the negative energy states exploitable. 
Importantly,
negative energy spin waves are the common ingredient for the realization of black-hole horizon analogues with spin waves and spin-wave superradiance. In this paper we show how to implement both. For this, we consider two set-ups.

The first set-up for the amplification of spin waves, consists of an antiferromagnet (AFM) subject to an external magnetic
field that varies in space. On one side of the AFM the field is smaller than the spin flop field, and on the other side it is larger. When the field is smaller than the spin-flop field, the system is energetically stable and the two sublattices in the AFM are antiparallel. The spins in the part of the AFM where the field is bigger than the spin-flop field, would normally reorient toward the spin-flop phase in which the spins are no longer fully antiparallel. This reorientation, is, however, prevented by spin-orbit torque. As a result the system supports negative energy spin waves that, upon scattering with spin waves from the region where the field is smaller than the spin flop field, lead to the enhanced reflection.  

This enhanced reflection of spin waves is an example of superradiance, namely, the amplitude of the reflected
wave is bigger than the amplitude of the incident
wave. Superradiance is a classical phenomenon that occurs
in several systems, such as scattering of electromagnetic waves off a rotating conducting cylinder~\cite{cylinder,lahaie1979scattering,ho2009simulation}, surface waves in water scattering off a draining bathtub~\cite{torres2017rotational}, and objects breaking apart in the ergosphere of a rotating black hole (Penrose’s
process~\cite{penrose1971extraction,finster2019lectures,sciama1975physics}).
Superradiance is based on the existence of energy stored in some degree of freedom of the system, which provides the extra energy needed for the amplification.

In ferromagnets, amplification of spin waves through the Klein paradox~\cite{Kleinparadox,manogue1988klein,calogeracos1999history,dombey1999seventy,krekora2004klein} was proposed theoretically by Harms {\it et  al.}~\cite{PhysRevApplied.18.064026}, to enable propagation of spin waves over longer distances.
At present we choose as a platform AFM because of the advantage of fast magnetization dynamics.
While writing this article we became aware of a recent paper on AFMs~\cite{ShaohuaYuan}. The difference between this latter work and ours is that we use SOT to stabilize the set-up such that the spin-wave amplification is a steady-state phenomenon rather than a transient effect.

The second set-up we propose consists of an analogue black-hole horizon for antiferromagnetic spin waves and will be useful for fundamental research. In fact analogues of horizons were theoretically proposed in 2017~\cite{roldan2017magnonic} in ferromagnets, exploiting STT. The STT, or, rather, spatial variations of STT, simulates gravitational fields for spin waves. This set-up is particularly relevant to investigate some features of black hole horizons because of the linear dispersion of AFM in case of zero anisotropy.
In particular, we propose to implement the horizon with a spatially varying exchange interaction in the AFM. We find the critical current and the values of the parameters needed to implement the horizon. 
This set-up is particularly attractive to investigate Hawking radiation of spin waves since the Hawking temperature scales directly with the velocity of the waves which is larger for antiferromagnets than for ferromagnets.

In the remainder of this article we first introduce our model. 
Subsequently, we compute the spin wave dispersion relation and we describe positive and negative energy excitations. In Sec. III we include two pumping mechanisms to inject angular momentum, i.e., SOT and STT. In Sec. IV we describe how to combine these elements to have the desired effect of spin wave-amplification with SOT and how to implement horizons with STT. We compute the corresponding critical currents to achieve spin-wave amplification and analogue horizons. We quantify the spin-wave amplification with an effective Klein-Gordon equation. We end in Sec. V with a conclusion, general discussion, and outlook.
\begin{figure}[h!]
\centering
\includegraphics[width=8.5cm]{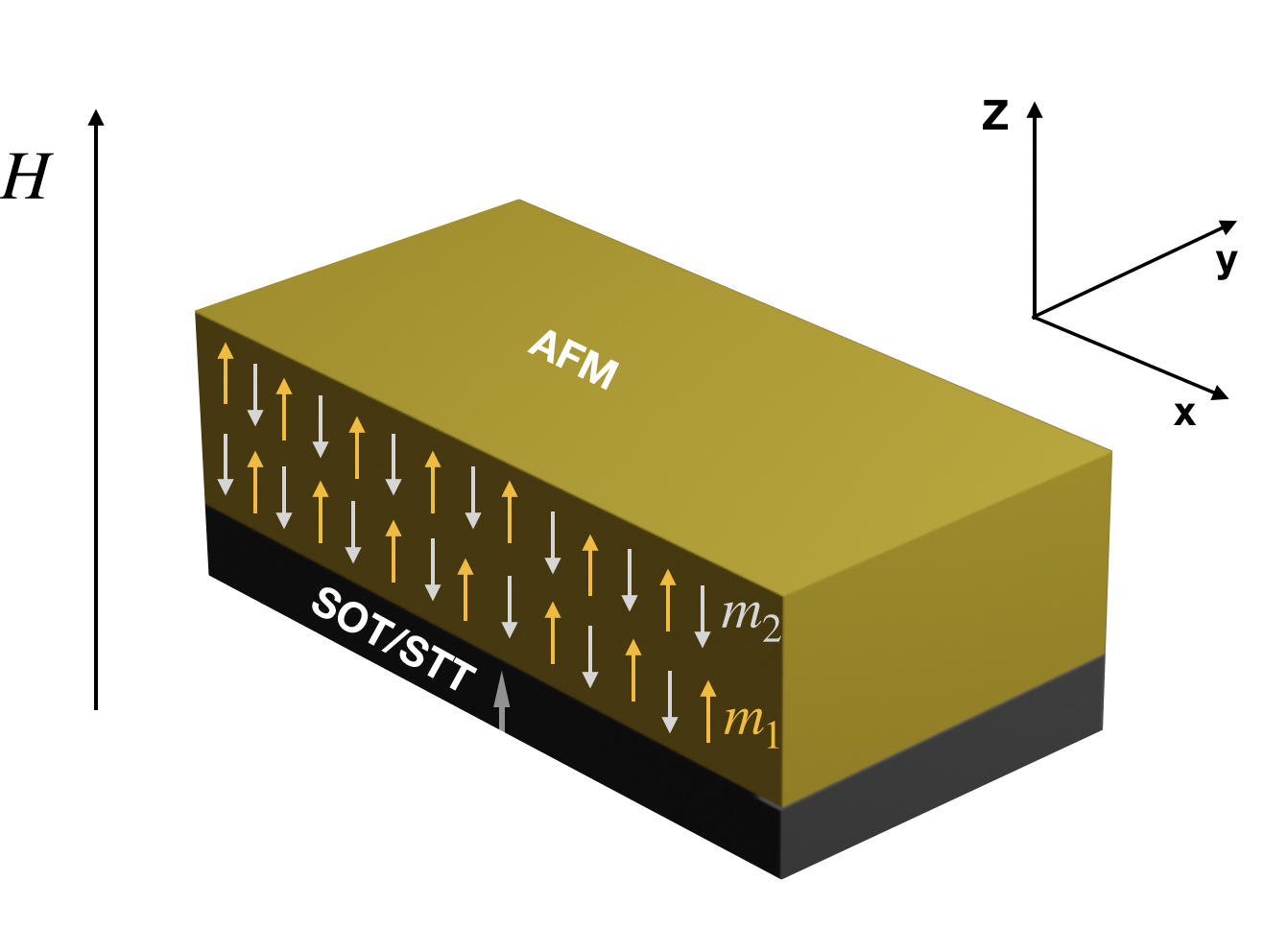}
\caption{An antiferromagnet in which the two sublattices $\mathbf{m}_1$ and $\mathbf{m}_2$ are indicated. A torque is exerted on the magnetization dynamics through spin orbit torque (SOT) or spin transfer torque (STT). The system is subject to an external magnetic field $\mathbf{H}$ oriented along the $\hat{z}$ direction.}
\label{fig:my_label}
\end{figure}

\section{Positive and negative energy spin-waves}
In this section we introduce the model, and discuss the spin-wave dispersion relations.
\subsection{The model}
We consider a collinear tetragonal antiferromagnet~\cite{bogdanov2002magnetic}, subject to an external field and SOT or STT (see Fig.~1). Within a continuum description consistent with the symmetries and the two-sublattice
magnetic structure, the functional of the magnetizations $\mathbf{m}_1$ and $\mathbf{m}_2$ on the two sublattices is as follows~\cite{bogdanov2002magnetic}:
\begin{align}
W&=\int \left\{\frac{\alpha}{2} \sum_{j=1,2,3} \left[ \left(\frac{\partial \mathbf{m}_1}{\partial x_j}    \right)^2+ \left(\frac{\partial \mathbf{m}_2}{\partial x_j}    \right)^2     \right]\right. \nonumber \\
&\left.+\alpha' \sum_{j=1,2,3}\left( \frac{\partial \mathbf{m}_1}{\partial x_j} \frac{\partial \mathbf{m}_2}{\partial x_j}\right)
 +\frac{\lambda}{2} \mathbf{m}_1 \cdot \mathbf{m}_2-\mathbf{H} \cdot (\mathbf{m}_1+\mathbf{m}_2)\right. \nonumber \\
  &\left.-\frac{\beta}{2}(m_{1z}^2+m_{2z}^2)-\beta' m_{1z}m_{2z}   \right\} dV .
\label{eqn:EnergyFunctional}
\end{align}
 The inhomogeneous and homogeneous part of the exchange coupling are expressed, respectively, in the $(\alpha,\alpha')$ and $\lambda$ terms. The next term describes the interaction with the external field $\mathbf{H}$, while the last two terms give rise to uniaxial second order magnetocrystalline anisotropy. The $z$-axis is
along the tetragonal axis of the antiferromagnet and from now on we consider the magnetic field oriented along this axis, i.e. $\mathbf{H}=(0,0,H)$, with $H>0$.

This magnetic energy is valid for temperatures sufficiently far below the Curie
temperature. In this regime, the vectors of sublattice magnetization $\mathbf{M}_\gamma$ $(\gamma=1,2)$ do not change their modulus, and they have only orientational degrees
of freedom. Therefore the antiferromagnet is described by the unit vectors
$\mathbf{m}_\gamma = \mathbf{M}_\gamma/M_s$, where $M_s = |M_1| = |M_2|$ is the sublattice
saturation magnetization. We use the sublattice approach instead of introducing the Néel vector and total magnetization to allow for straightforward quantization in future work.
\subsection{Equations of motion and dispersion relation}
\begin{figure}[h!]
    \centering
    \includegraphics[width=8cm]{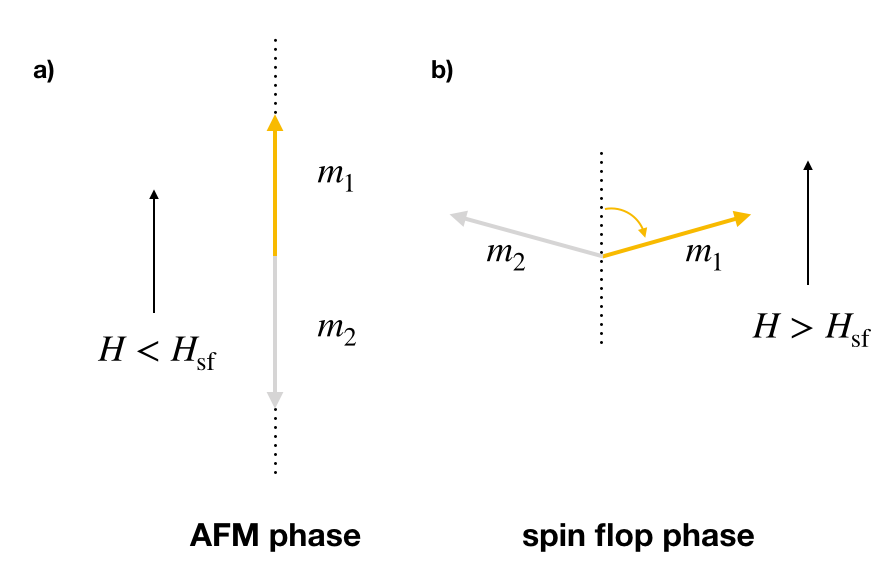}
    \caption{a) The orientation of the two sublattices $\mathbf{m}_1$ and $\mathbf{m}_2$ is antiparallel if the external magnetic field $H$ is smaller than the spin flop field $H_{\rm sf}$. The net magnetization $\mathbf{m}=\mathbf{m}_1+\mathbf{m}_2$ is then zero. b) In case the external magnetic field is larger than the spin flop field $H_{\rm sf}$, $\mathbf{m}_1$ and $\mathbf{m}_2$ "flop" and there is a net magnetization different from zero. }
    \label{fig:my_label}
\end{figure} 
In this subsection we compute the dispersion relation from the linearization of the equations of motion. We first discuss around which state we are linearizing. In order for our system to have negative energy states we should be in a stationary state, which is not the true ground state of the system. We first consider the antiferromagnetic configuration in Fig.~2a. 
For small fields, this configuration is the ground state of the system. 
Now, if we increase the magnetic field until and above the spin flop field $H_{\rm sf}$, the system obtains a nonzero net magnetization, $\mathbf{m}_1+\mathbf{m}_2>0$ (Fig.~2b). Therefore the configuration in Fig.~2b becomes the ground state of the system for magnetic fields bigger than $H_{\rm sf}$ (and much smaller than $\lambda$, above which the system becomes ferromagnetically ordered).\\
In the next section we describe how to stabilize the antiferromagnetic state even when $H>H_{\rm sf}$. Because of the stabilizing mechanism, the system is energetically unstable but dynamically stable. Therefore, we linearize the equations of motion around the AFM state for both $H<H_{\rm sf}$ and $H>H_{\rm sf}$.

The equations of motion for $\mathbf{m}_1$ and $\mathbf{m}_2$ are given by the Landau–Lifshitz equation
$$\frac{\partial \mathbf{{m}}_1}{\partial t}=-|\gamma| (\mathbf{m}_1\times \mathbf{H}_1);$$
$$\frac{\partial \mathbf{{m}}_2}{\partial t}=-|\gamma| (\mathbf{m}_2\times \mathbf{H}_2),$$
where $\mathbf{H}_\gamma \equiv -\delta w/\delta \mathbf{M}_\gamma $ are the effective magnetic fields, with $w= \delta W / \delta V$ the free energy (per unit volume) and
$|\gamma|$ the modulus of the gyromagnetic ratio. We substitute $\mathbf{m}_1=(
\delta m_1^x,
\delta m_1^y,
1)$, $\mathbf{m}_2=(
\delta m_2^x,
\delta m_2^y,
-1)$ in the equations of motion and keep terms up to first order in $\delta m$. We introduce the complex field $\psi=(\delta m_x+i \delta m_y)$, $\psi^*=(\delta  m_x-i \delta m_y)$ and rewrite the linearized equations of motion in the matrix form
\begin{equation}
i \frac{\partial }{\partial t}\begin{pmatrix}
\psi_1 \\  \psi_1^*\\
\psi_2 \\  \psi_2^*
\end{pmatrix}=
\mathscr{L}
\begin{pmatrix}
\psi_1 \\  \psi_1^*\\
\psi_2 \\  \psi_2^*
\end{pmatrix},
\label{eqn:Lequation}
\end{equation}
where $\mathscr{L}$ is given by
\begin{widetext}
\begin{equation}
\mathscr{L}=
\begin{pmatrix}
-\frac{\lambda}{2}-H-\beta+\beta'+\alpha \nabla^2  &0&-\frac{\lambda}{2}+\alpha'\nabla^2      &0\\ 
     0&  \frac{\lambda}{2}+H+\beta-\beta'-\alpha \nabla^2        &0&    \frac{\lambda}{2}-\alpha'\nabla^2       \\
\frac{\lambda}{2}-\alpha'\nabla^2    & 0 &\frac{\lambda}{2}-H+\beta-\beta'-\alpha \nabla^2   &0 \\  
0& -\frac{\lambda}{2}+\alpha'\nabla^2 & 0 & -\frac{\lambda}{2}+H-\beta+\beta'+\alpha \nabla^2 
\end{pmatrix}.
\end{equation}
\end{widetext}
The $2\times2$ block in the upper left and bottom right of the matrix refer respectively to the first and second sublattice, with the terms due to interactions between the two sublattices in the top left and bottom right $2 \times 2$ blocks. We notice that every time $\beta$ appears it is accompanied by $\beta'$, behaving, effectively, like an anisotropy term. We define $(\alpha-\alpha')\equiv A$ and $(\beta-\beta')\equiv K_z$. With the Bogoliubov ansatz ${\psi_\gamma}(\mathbf{x},t)=u_\gamma(\mathbf{x}) e^{-i \omega t}+v_\gamma^*(\mathbf{x})e^{i \omega^* t}$(with $u_\gamma,v_\gamma$ proportional to $e^{i k x }$), we find the dispersion relations. 
To simplify the dispersion relations it is important to recall the hierarchy of the interactions, i.e.  $\lambda \gg K_z$. Moreover, we consider the long wavelength limit $A k^2\ll \lambda$. Within those approximations, we obtain the dispersion relations \cite{Kosevich1981,Kim2014c}
\begin{equation}
\omega^0_{ \pm \pm}(k)= \pm \frac{H |\gamma|}{M_s} \pm  \sqrt{\left(\frac{K_z |\gamma|}{M_s}+\frac{A |\gamma|}{M_s} k^2\right)  \frac{\lambda |\gamma|}{M_s}},
\label{eqn:disprel}
\end{equation}
where the first and second subscript refers, respectively, to the sign in front of the magnetic field $H$ and to the sign in front of the square root. The superscript $0$ is for future convenience, to distinguish these frequencies without damping and pumping from the ones computed in the next section. In Fig.~3a we plot the four dispersions for $H<H_{\rm sf}=\sqrt{K_z \lambda}$.

Let us now discuss the relations between those four dispersions and corresponding modes. 
For generic eigenvalues $\omega_i^0$ with eigenvector $(u_{1i},v_{1i},u_{2i},v_{2i})$, we have
$$\mathscr{L}\begin{pmatrix}
    u_{1i} \\ v_{1i}\\ u_{2i} \\v_{2i}
\end{pmatrix}= \omega_i^0 \begin{pmatrix} u_{1i} \\ v_{1i}\\ u_{2i} \\v_{2i} \end{pmatrix}$$ 
where $i = 1, 2, 3, 4$ labels the eigenvalues and
eigenvectors of $\mathscr{L}$. 
If we find a $\tilde{\sigma_1}$, such that $\tilde{\sigma_1}\mathscr{L}\tilde{\sigma_1}=-\mathscr{L}^* $, one can show that it follows
$$\mathscr{L}\begin{pmatrix}
    v_{1i}^* \\ u_{1i}^* \\v_{2i}^* \\ u_{2i}^*\end{pmatrix}= -(\omega^0_i)^* \begin{pmatrix}  v_{1i}^* \\ u_{1i}^* \\v_{2i}^* \\ u_{2i}^*\end{pmatrix}.$$
The desired $\tilde{\sigma_1}=\tilde{\sigma}_1^{-1}$ is
$$\tilde{\sigma}_1= \begin{pmatrix} 0 & 1 &0 &0 \\
1 & 0 & 0 & 0\\
 0 & 0 & 0 & 1 \\
 0 & 0 & 1 & 0
\end{pmatrix}.$$
\begin{figure}
    \centering
    \includegraphics[width=8cm]{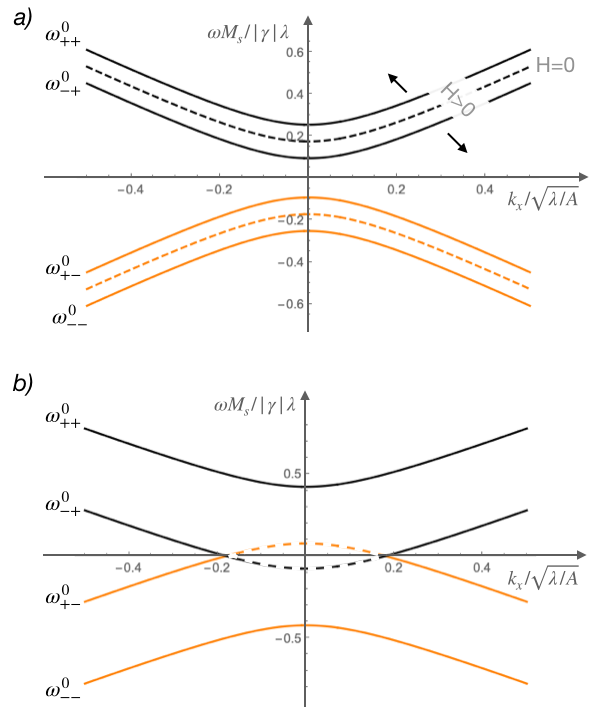 }
    \caption{Plot of the spin-wave dispersion relation \cite{Kim2014c}. In black are the positive norm modes and in orange the negative norm modes. The external magnetic field lifts the degeneracy. In a) the magnetic field is smaller than the spin flop field, i.e. $H<H_{\rm sf}$. For increasing field, the lowest (highest) positive (negative)-norm mode goes down (up) and becomes zero at $k=0$ for $H=H_{\rm sf}$. b) Increasing further the magnetic field, such that it becomes bigger than $H_{\rm sf}$, $\omega_{-+}^0$ ($\omega_{+-}^0$) becomes negative (positive) for a certain interval of $k$. The negative energy modes are indicated with dashed lines. The values for the plot are $K_z/ \lambda=0.03$, a) $H/\lambda=0.08$, b) $H/\lambda=0.25$. }
    \label{fig:my_label}
\end{figure}
For the four dispersions in Eq.~(\ref{eqn:disprel}), we confirm that if $\omega^0_i$ is an eigenfrequency also $-(\omega^0_i)^*$ is an eigenfrequency. That is $\omega^0_{++}=-\omega^0_{--}$ and $\omega^0_{+-} =-\omega^0_{-+}$.
Hence, the spin wave dispersion relation come in pairs that are related by $\omega^0_i= -(\omega^0_i)^*$. This doubling is the result of introducing the complex spinor $\psi$. As we discuss next, these two modes, that physically correspond to the same excitation, have opposite norm. 
\subsection{Negative energy modes}
To find the energy contribution of the modes, we expand the energy functional $W$ in Eq.~(\ref{eqn:EnergyFunctional}) in terms of increasing order in the fluctuations $\delta m$, such that $W = W^{(0)} +W^{(1)} +W^{(2)}$, where the superscript indicates the order in the fluctuations. The first-order term vanishes because we linearize around a (meta)stable state. The second-order term is the first contribution to the energy due to the fluctuations. In terms of the complex field $\psi$, we obtain
$$W^{(2)}=\frac{1}{4} \langle \psi |M|  \psi \rangle,$$
where $M=\tilde{\sigma}_3\mathscr{L}$ and  $|\psi \rangle =(\psi_1,\psi_1^*, \psi_2, \psi_2^*)^T$. Expanding the field in terms of the eigenfunctions, i.e. $|\psi\rangle= \sum_i a_i |\phi_i \rangle $, we find
\begin{equation}
    W^{(2)}= \sum_{i=1,2,3,4} |a_i|^2 \langle \phi_i | \tilde{\sigma_3} | \phi_i \rangle \omega_i,
    \label{eqn:multiplication}
    \end{equation}
where $|\phi_i \rangle$ is the eigenvector with components $|\phi_i \rangle = u_{1i},v_{1i},u_{2i},v_{2i})^T$ and 
$$
\tilde{\sigma_3}=\frac{1}{2}\begin{pmatrix}
-1  &0&  0  &0\\ 
     0&  1      &0& 0        \\
 0 & 0 &1   &0 \\  
0& 0 & 0 & -1
\end{pmatrix}
.$$
In doing those calculations we considered the argument under the square root positive. We now introduce the conserved norm of $\phi_i$ as
\begin{equation}
    \langle \phi_i| \tilde{\sigma_3}| \phi_i \rangle=(-u_{1i}^*u_{1i}+v_{1i}^*v_{1i}+u_{2i}^*u_{2i}-v_{i2}^*v_{2i}).
    \label{eqn:norm}
    \end{equation}
This norm is conserved since $\tilde{\sigma}_3\mathscr{L}\tilde{\sigma}_3=\mathscr{L}$.
We also notice that $\mathscr{L}$ is a pseudo-hermitian matrix, i.e.,  $\tilde{\sigma_3}^{-1}\mathscr{L}^{\dagger} \tilde{\sigma_3}=\mathscr{L}$.

For the dispersion in Eq.~(\ref{eqn:disprel}) we find that $\omega_{++}^0$ and $\omega_{-+}^0$ have positive norm, and that $\omega_{--}^0$ and $\omega_{+-}^0$ have negative norm.
From Eq.~(\ref{eqn:multiplication}) we observe that the sign of the norm multiplied by the corresponding frequency tells us whether the contribution to the energy of that particular excitation is positive or negative. 
In fact, as mentioned before, the pair of frequencies $\omega_{--}^0$ and $\omega_{++}^0$, and the pair $\omega_{+-}^0$ and $\omega_{-+}^0$ physically correspond to the same excitation. Specifically, from Eq.~(\ref{eqn:disprel}), the eigenfrequencies $\omega^0_{++}$ and $\omega_{--}^0$ are respectively positive and negative for every momentum $k$, while $\omega^0_{+-}$ is negative for $H<H_{\rm sf}$ and $\omega^0_{-+}$ is positive for $H<H_{\rm sf}$.
As it is possible to see in Fig.~3b, $\omega_{+-}^0$ can become a negative energy mode when $H>H_{ \rm sf}$. Hence, for fields larger than the spin flop field the modes $\omega^0_{+-}$ and $\omega^0_{-+}$ correspond to negative energy excitations over a range of wave vectors (see Fig.~3). The negative-energy excitations result from the linearization of the magnetization dynamics around the antiferromagnetic state for fields larger than the spin-flop field, i.e., from the linearization of the dynamics around a state that is not the true ground state. Hence, the system is energetically unstable and in the presence of dissipation would dynamically transition from the antiferromagnetic to the spin-flop state. For our purposes, we would like to use the negative-energy excitations and hence we need to dynamically stabilize the system. Therefore the next section is dedicated to the discussion of this stabilization.
\section{Dynamic stabilization}
In this section we describe the results of exerting a torque on the magnetization through spin orbit torque and spin transfer torque~\cite{fukami2020antiferromagnetic,hals2011phenomenology,gomonay2014spintronics,gomonay2010spin,cheng2014spin,xu2008spin,ryu2020current,vzelezny2018spin,shao2021roadmap,johansen2018spin,brataas2020spin}. We will discuss how the dispersion relation changes and how to dynamically stabilize the system, making the presence of long-lived negative energy excitations possible.
\subsection{Spin-orbit torques}
For dynamic stabilization by spin-orbit torques, we consider a current passing through a heavy metal layer, on top of an AFM insulator. The microscopic mechanism is that, thanks to the large spin orbit coupling typical of the heavy metal the charge current is converted into spin current, and there is a 
spin accumulation in the heavy metal at the interface.
The net result is that a torque is exerted on the two sublattices of the AFM insulator. 
The equations that phenomenologically describe the dynamics are~\cite{gomonay2010spin}
$$\mathbf{\dot{m}}_j=-\gamma (\mathbf{m}_j\times \mathbf{H}_j)+I_{s1} \mathbf{m}_j \times(  \mathbf{m}_j \times \hat{z} )$$ $$+I_{s2} \mathbf{m}_j \times ( \mathbf{m}_l\times  \hat{z}) +\alpha_0 \mathbf{m}_j \times \dot{\mathbf{m}}_j + \alpha_1 \mathbf{m}_j \times \dot{\mathbf{m}}_l, $$
where $j \neq l$ is the index of the sublattice and the terms with the coefficients $I_{s1},I_{s2}$ account for the spin orbit torques. We have also added a intrasublattice Gilbert damping $\alpha_0$ and an intersublattice Gilbert damping $\alpha_1$. In the equations of motion the terms describing torques are symmetric under exchange of the two sublattices, in the same way as the damping terms are.
In the long wavelength limit, for $\lambda \gg K_z, A k^2$ and up to first order in $\alpha \equiv (\alpha_0-\alpha_1)/ 2$ and $I_s \equiv (I_{s1}-I_{s2})/2$, we find the dispersion relations 
\begin{equation}
\omega_{\pm -}(k) \simeq \omega_{\pm -}^0-i   \alpha \left( \frac{\lambda |\gamma|}{ M_s}  \mp \frac{H |\gamma|}{M_s} f(k) \right) \mp  i I_s f(k),
\end{equation}
\begin{equation}
\omega_{\pm +}(k) \simeq \omega_{\pm +}^0-i   \alpha \left( \frac{\lambda |\gamma|}{ M_s}  \pm \frac{H |\gamma|}{M_s} f(k) \right) \pm  i I_s f(k).
\end{equation}
where $f(k) = \sqrt{\lambda/(2 K_z)} - (A \sqrt{\lambda})/( 2^{3/2} K_z^{3/2}) k^2$.

We notice that the real part of the dispersion relation coincides with $\omega^0_{\pm \pm}$, plotted in Fig.~3.
We also notice that the ${\sigma}_1$ symmetry is respected, i.e., for every frequency $\omega_i$ there is a frequency $-\omega_i^*$.
Referring to our result in the previous section for negative energy excitations, we know that in order to have negative energy excitations, we need $H>H_{\rm sf}$. For dynamic stability we need, on the other hand, that the imaginary part of the frequency is negative. This results in the requirement $I_s>\alpha (H-H_{\rm sf})|\gamma|/M_s$. Note that it is possible to change the sign of $I_{s}$ by changing the direction of polarization of the current or changing the direction of the current.
\begin{figure}[h!]
    \centering
    \includegraphics[width=9cm]{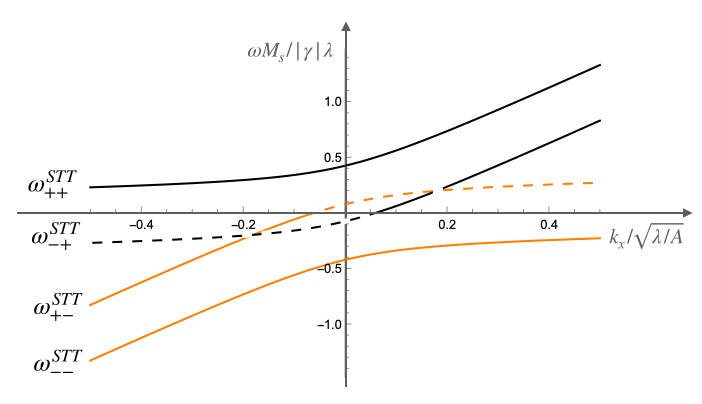 }
    \caption{Plot of the real part of dispersion relation with STT for $H>H_{\rm sf}$. Positive norm modes are indicated in black and negative norm modes in orange. Negative energy modes are indicated with dashed lines. The values for the plot are $H/\lambda=0.25$, $K_z/ \lambda=0.03$, $\frac{v M_s}{|\gamma| \sqrt{A \lambda}}=1.1$. }
    \label{fig:my_label}
\end{figure} 
\subsection{Spin-transfer torque}
For the purpose of dynamic stabilization by spin-transfer torques we consider a charge current flowing in a conducting AFM. The spin of the conduction electrons interacts with the local magnetic moments and this gives rise to a torque on the two sublattices.
This is phenomenologically described by the equations~\cite{manchon2019current,hals2011phenomenology}
$$\left( \frac{\partial}{\partial t} +\mathbf{v} \cdot \nabla  \right)\mathbf{{m}}_j=-\gamma (\mathbf{m}_j\times \mathbf{H}_j)$$ 
$$+\mathbf{m}_j \times  \left(\alpha_0 \frac{\partial}{\partial t} +\beta_0 \mathbf{v} \cdot \nabla  \right) {\mathbf{m}}_j + \mathbf{m}_j \times  \left(\alpha_1 \frac{\partial}{\partial t} +\beta_1 \mathbf{v} \cdot \nabla  \right)    {\mathbf{m}}_l .$$
The terms containing the velocity $\mathbf{v}$ represent the reactive and dissipative spin transfer torque, the dimensionless parameters $\beta_0,\beta_1 $ being the dissipative coefficients. 
For concretness, we consider $\mathbf{v}$ along the $x$-direction, i.e. $\mathbf{v}=(v,0,0)$.
Following the same procedure as in the second section, we compute the corresponding dispersion relations. In the long wavelengh limit, for $\lambda \gg K_z, A k^2$ and up to first order in $\alpha \equiv (\alpha_0-\alpha_1)/ 2$ and $\beta \equiv (\beta_0-\beta_1)/2$, we find
$${\rm Re}[\omega_{\pm \pm}]= \omega_{\pm \pm}^0+k_x v,$$
\begin{equation}
\begin{split}
{\rm Im}[\omega_{\pm -}]=-\alpha \left( \frac{\lambda |\gamma|}{ M_s} \mp f(k)\left( \frac{H |\gamma|}{M_s} \pm  k_x v \right)  \right)-\beta f(k) k_x v,
\end{split}
\end{equation}
\begin{equation}
\begin{split}
{\rm Im}[\omega_{\pm +}]=  - \alpha\left( \frac{\lambda |\gamma|}{ M_s} \pm f(k) \left( \frac{H |\gamma| }{M_s} \pm  k_x v \right) \right) +  \beta f(k)  k_x v .
\end{split}
\label{eqn: STT}
\end{equation}
where $f(k) = \sqrt{\lambda/(2 K_z)} - (A \sqrt{\lambda})/( 2^{3/2} K_z^{3/2}) k^2$.

Looking at the real part of the dispersion relation, we notice that the spin-transfer torque gives the spin waves a Doppler shift $k_x v$ (see Fig. 4). In other words, the spin waves are dragged along with the velocity $\mathbf{v}$ as a consequence of the spin current. 
On the other hand, we notice that the dissipative spin transfer torque can compensate the damping and contribute to the dynamical stability of the system. 
 
The existence of negative energy modes requires $H>(\lambda \sqrt{A K_z}-(v M_s/|\gamma|)^2 \sqrt{K_z/A})/\sqrt{A \lambda- (v M_s/|\gamma|)^2}$ for $0<v<\sqrt{A \lambda}|\gamma|/M_s$ and $H>(\lambda \sqrt{A K_z}+(v M_s/|\gamma|)^2 \sqrt{K_z/A})/\sqrt{A \lambda- (v M_s/|\gamma|)^2}$ for $-\sqrt{A \lambda}|\gamma|/M_s<v< 0$.
To achieve energetic instability for all positive $k_x$, in addition to the magnetic field being bigger than the spin flop field, the condition for the velocity is $v>\sqrt{A \lambda}|\gamma|/M_s$, while for all negative $k_x$ the velocity has to be $v<-\sqrt{A \lambda}|\gamma|/M_s$. To dynamically stabilize this system it is enough for $\beta$ to be equal to $\alpha$, or, more generally, $-\sqrt{A \lambda}< v(\beta/ \alpha -1)M_s/|\gamma|< \sqrt{A \lambda}$.
\section{Applications and discussion }
\begin{figure*}
    \centering
    \includegraphics[width=0.85\textwidth]{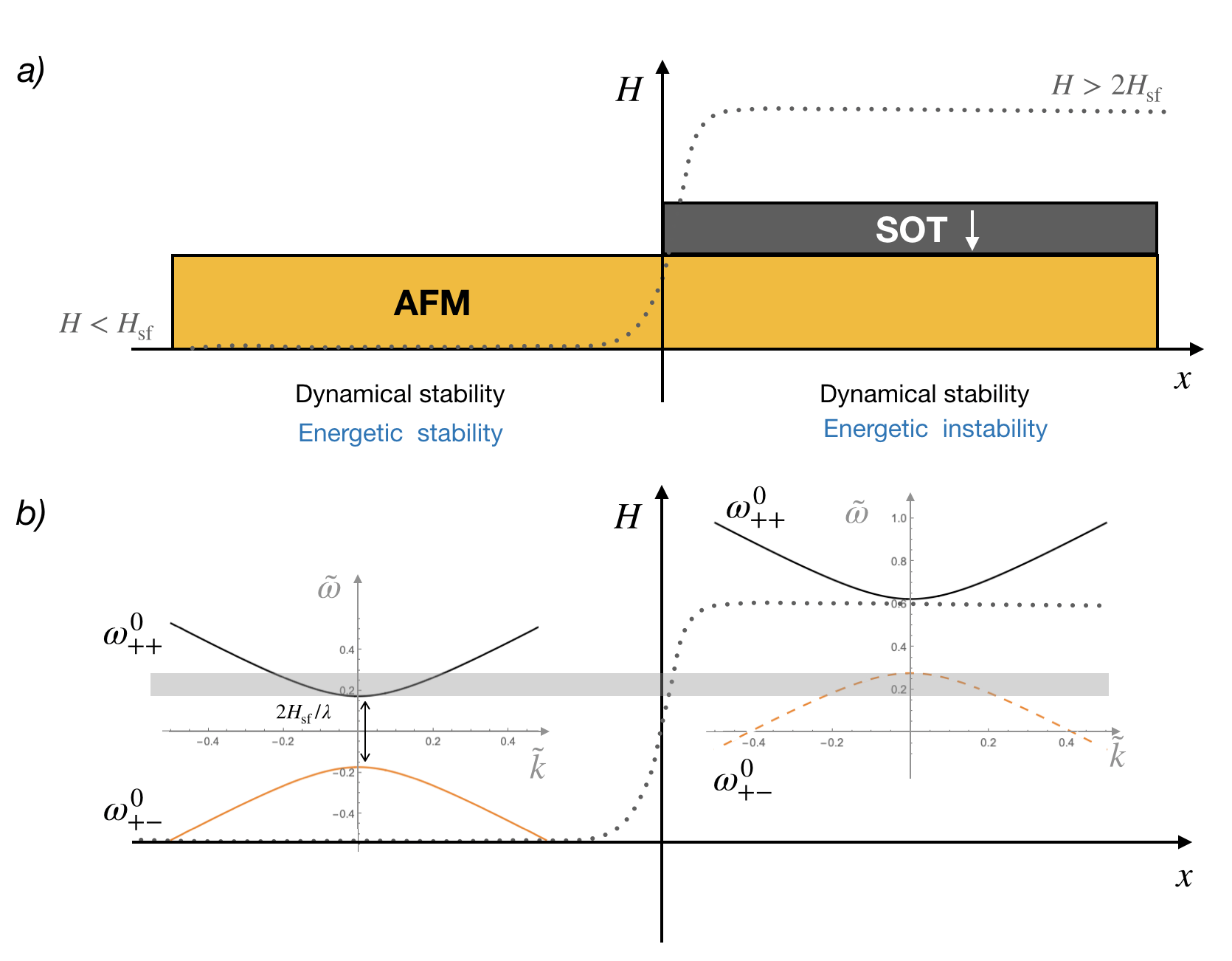  }
    \caption{Klein paradox: enhanced reflection and therefore amplification of spin waves. a) Set-up proposed for achieving amplification of spin waves. An antiferromagnet in part of which angular momentum is injected throught SOT. The external magnetic field varying in space is indicated by the dotted line. b) Corresponding dispersion relations, with $H>2 H_{\rm sf}$ in the right part. For the interval of frequencies indicated by the grey horizontal region, there is an overlap of the positive norm band (black) in the left part of the AFM with the negative norm band (orange) in the right part. This overlap is tuned by the inhomogeneous external field $H$ which has the effect of vertically shifting the dispersion relation. For the interval of frequencies indicated in grey amplification, i.e. superradiant scattering, is possible. In the graph the dimensionless quantities $\tilde{\omega}=\omega M_s/|\gamma|\lambda$, $\tilde{k}= k (   \sqrt{A / \lambda})$. The values for the plot are  $K_z/ \lambda=0.03$ and $\bar{H}=0.45 \lambda$.}
    \label{fig:img1}
\end{figure*}

\begin{figure}
    \centering
    \includegraphics[width=8.5cm]{ 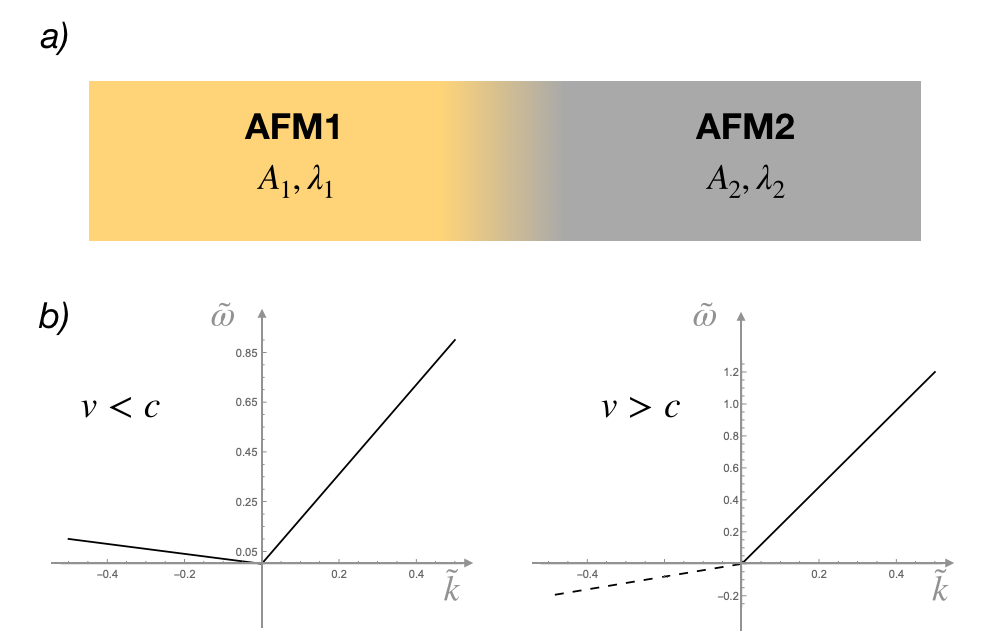 }
    \caption{a) Set-up for creating a sonic horizon for spin waves incoming from left. The background velocity is implemented thanks to STT. The background velocity $\mathbf{v}\propto \mathbf{j}$ is varying in space, which for example can be realized through shrinking~\cite{roldan2017magnonic} or with different exchange in the right part respect to the left, as schematized.
   b) Corresponding dispersion relations of the right and left part for $K_z=H=0$. With this particular choice of parameters, this set-up is particularly useful as analogue of black holes horizons thanks to the linear dispersion relation. 
   In the left dispersion relation $v<c$, while in the right one $v>c$, i.e. for background velocities bigger than the sound velocity $c=\sqrt{A \lambda}$ spin waves are forced to move toward right. With dashed line negative energy mode is indicated.
   In the graph the dimensionless axis are $\tilde{\omega}=\frac{\omega}{|\gamma|\lambda/M_s}$ and $\tilde{k}= k \sqrt{A / \lambda}$. The values for the plot are $H=0$, $K_z=0$, b)$\frac{v M_s}{|\gamma |\sqrt{A \lambda}}=0.8$ (left), $\frac{v M_s}{|\gamma| \sqrt{A \lambda}}=1.4$ (right).   }
    \label{fig:img1}
\end{figure}
In this section we are going to discuss two possible applications of the dynamically stable negative-energy spin waves: i) the amplification of spin waves and ii) the implementation of an event horizon for spin waves.
\subsection{Klein paradox: enhanced reflection, i.e amplification of spin waves}
Starting from the set-up in Fig.~1 it is possible to built a set-up for amplification of spin waves. It consists of the following. One part of the antiferromagnet is subject to a torque due to a current flowing in the heavy metal (Fig.~5a). The external magnetic field is varying in space as indicated with the dotted lines and in the right part it reaches a value bigger than $H_{\rm sf}$. As a consequence, the torque is applied on the right part of the antiferromagnet in order to keep the total configuration of the two sublattices antiparallel, ensuring dynamical stability. We assume that the AFM state is homogeneous in all the AFM. The dispersion relations for $x \rightarrow \pm \infty$ of the right and left parts are indicated in Fig.~5b, where we plot only $\omega_{++}$ and $\omega_{+-}$. 

For an interval of frequencies (in grey) there is overlap of the positive norm band (black) in the left part of the AFM that correspond to positive energy excitations, with the negative norm band (orange) in the right part, that correspond to negative-energy excitations. For those frequencies superradiant scattering is possible due to the coupling of positive-energy excitations on the left with negative energy excitations on the right.

For large exchange $\lambda$, an effective description of the spin-wave excitations in an antiferromagnet follows from~\cite{shen2020driving}
\begin{equation}
-K_z\varphi+A \frac{\partial^2 \varphi}{\partial x^2}-\frac{1}{ \lambda} \left(  \frac{M_s}{|\gamma|}\partial_t+ i  H  \right)^2\varphi=0,
\label{eqn: KleinAFM}
\end{equation}
where $\varphi= \mathbf{\delta n } \cdot \hat{x}-i \mathbf{\delta n} \cdot \hat{y}$, with $\mathbf{n}= (\mathbf{m}_1-\mathbf{m}_2 )/2$ the Neel vector and $\mathbf{\delta n}$ a small deviation orthogonal to $\mathbf{n}_0=\mathbf{\hat{z}}$.

For our purposes, we consider a quasi-one-dimensional set-up such that we only take into account the $x$-dependence of the magnetization dynamics. Inserting plane wave ansatz, this equation yields the frequency $\omega^0_{+\pm}$.
This equation is equivalent to the Klein-Gordon equation for a charged scalar field $\phi$ in one spatial dimension coupled to a spatially-varying electrostatic potential $A(x)$~\cite{hund1941materieerzeugung}
$$ -\left( \frac{1}{c} \partial_t +i \frac{e}{\hbar c}A\right)^2 \phi +\partial_x^2 \phi -\frac{\mu^2 c^2}{\hbar^2} \phi=0,$$
where $c$ is the speed of light, $e$ is the electric charge of the field, $\mu$ its mass and we take the electrostatic potential asymptotically constant with $A(x \rightarrow-\infty) = 0$ and $A(x \rightarrow +\infty) = \bar{A}$.
We notice that the external magnetic field plays the role of the electrostatic potential and the mass of the field arises from the anisotropy.

In the original Klein paradox, superradiant reflection occurs if the electrostatic potential step is larger than twice the restmass~\cite{giacomelli2021superradiant}, which leads to the enhanced scattering, is then possible.

Similarly, from Fig.~5, it is easy to understand that superradiant scattering  occurs only if the magnetic field is strong enough to overcome the frequency gap
$2 H_{\rm sf} |\gamma|/M_s$. Here, we take the field such that $H=0$ for $x \to -\infty$ and $H=\bar{H}>2 H_{\rm sf}$ for $x \to +\infty$.
Superradiant scattering occurs for frequencies
$ H_{\rm sf} < M_s \omega / |\gamma| < \bar{H} - H_{\rm sf}$.
For $\bar{H}-H_{\rm sf} < M_s \omega / |\gamma| <  \bar{H}+H_{\rm sf}$ total reflection takes place as a consequence of the absence of modes for $ x \rightarrow +\infty$, while for higher frequencies conventional scattering occurs. For details see Appendix A. 

In the following we quantify the amplification. Proceeding in the same way as reference~\cite{giacomelli2021superradiant}, expanding in plane waves 
$\varphi(x,t)=\varphi(x)e^{-i \omega t}$ with
$$ \varphi(x)=
\begin{cases}
    = e^{i k_{-\infty}x}+ \mathcal{R} e^{-i k_{-\infty}x}\qquad \text{for}\quad  x \rightarrow -\infty ,\\
    =\mathcal{T} e^{i k_{+\infty}x}\qquad \qquad \qquad  \quad \text{for} \quad x \rightarrow + \infty\end{cases}
$$
where 
$A k_{+ \infty}^2=-K_z+(\bar{H}- \omega M_s / |\gamma|)^2/\lambda $
and $A k_{- \infty}^2=-K_z+( \omega M_s / |\gamma|)^2 / \lambda $,
we can write the relation between reflection $\mathcal{R}$ and transmission coefficient $\mathcal{T}$, i.e.,
$$ |\mathcal{R}|^2=1-\frac{k_{+\infty}}{k_{-\infty}}|\mathcal{T}|^2  .$$
In fact, if $k_{+\infty}/k_{-\infty}>0$, $|\mathcal{R}|^2<1$, i.e. conventional scattering occurs, while if $k_{+\infty}/k_{-\infty}<0$, then $|\mathcal{R}|^2>1$, meaning superradiant scattering. For details see Appendix B. As a result, a spin wave coming in from the left in Fig.~5 is amplified while reflecting back if it has the appropriate frequency.
\subsection{Event horizon for spin waves}
The second application is to engineer analogue black-hole horizons for spin waves in antiferromagnets. 
The set-up consists of a metallic antiferromagnet with negligible anistropy and with an external field in which the exchange varies in space, and through which a current flows, as in Fig~6a. The choice $K_z=0$ for this set-up is particularly
useful because it yields a linear spin wave dispersion and therefore Lorentz invariance at small wavelengths. 
In fact, the dispersion relation found in Sec.~IIIB (Eq.~$(\ref{eqn: STT})$) for the case $H=K_z=0$ becomes
$\omega_{\pm +} = |\gamma|\sqrt{A \lambda}|k| /M_s+ k_x v $.
As we saw in Sec.~III, the real part of the dispersion relation presents a drift term $k_x v$, which can be interpreted as Doppler shift, i.e, thanks to the torque from the current flowing in the AFM it is possible to implement a background velocity.

We introduce the spin wave speed $c=\sqrt{A \lambda}$. For simplicity we consider positive $v$. Two regimes are possible, one in which $v>c$ and the other in which $v<c$. This means that if the speed $c$ is lower than the
background velocity $\mathbf{v}=(v,0,0)$, spin waves are dragged and forced to travel to the
right. For clarity we plot the positive norm modes in Fig.~6b. In dasched line negative energy modes are shown. See Appendix C for details.

In this way, creating a spatial transition from $v<c$ to $v>c$ (Fig.~6a), implements a horizon for spin waves incoming from the left. A possibility to implement different spin wave velocities in the right part with respect to the left, is to spatially vary $A$ or $\lambda$ (Fig.~6a). Different exchange between two AFMs can be realized by chemical doping or by applying strain. The shading in Fig.~6a indicates the need to not have a sharp interface, that could make interface effects dominant. But varying the exchange arises the problem of the constant Néel temperature along all the AFM. The system should be such that the Néel temperature on both sides remains sufficiently large.

 Based on this proposal for engineering an analogue black-hole horizon for spin waves in antiferromagnets, one can study quantum effects such as Hawking radiation.

\section{Conclusions and outlook}
We have shown that negative-energy spin waves in AFMs are a versatile platform for both the laboratory study of high-energy physics and spin-wave amplification. In particular we have shown that it is possible to achieve spin-wave amplification in an antiferromagnet subjected to a spatially varying external magnetic field, in which torque is exerted on the magnetization dynamics thanks to a current flowing in a heavy metal adjacent to the AFM. If the asymptotic value of the magnetic field is bigger than twice the spin-flop field, it is possible to excite negative energy spin waves, enabling the amplification for a specific interval of frequencies. 
Antiferromagnets can also be used to implement event horizons for spin waves, in which the gravitational filed is caused by STT (as in Fig.~6). The horizon can be due to a varying background velocity $\mathbf{v}\propto \mathbf{j}$ or due to a spatially varying group velocity that results from varying the exchange constant. AFMs are particularly suitable platforms for analogue gravity because of the fast magnetization dynamics and the linear dispersion relation in case of small anisotropy.

We hope that our work and the ones of~\cite{ShaohuaYuan,adorno2023schwinger} will stimulate further research toward spin-wave amplification for energy-efficient devices based on AFMs and towards a further comprehension of analogue event horizons. As a further outlook it could interesting to study spin-wave amplification in altermagnets~\cite{vsmejkal2022emerging}.\\

R.A.D. is member of the D-ITP consortium, a program of
the Dutch Research Council (NWO), funded by the Dutch
Ministry of Education, Culture and Science (OCW). This
work is part of the research programme Fluid Spintronics
with project number 182.069, financed by the Dutch Research
Council (NWO).

\appendix
\section{Negative energy and positive energy modes with Lagrangian formalism}
To confirm results from section IIC (i.e. looking at Fig.~5b positive norm modes in black and negative norm modes in orange), we compute the norm from the Lagrangian formalism as well. 
In particular, starting from the Lagrangian of an AFM~\cite{shen2020driving} up to a constant
\begin{equation}
\begin{split}
\mathcal{L}= \frac{1}{\lambda} \left( \frac{M_s}{|\gamma|}\dot{\varphi}+i H \varphi\right)   \left( \frac{M_s}{|\gamma|}\dot{\varphi^*}-i H \varphi^*\right)-A(\partial_x \varphi) (\partial_x\varphi^*)\\-K_z \varphi \varphi^*
\end{split}
\end{equation}
we define the scalar product~\cite{ robertson2012theory,birrell1984quantum}
$$(\varphi_1, \varphi_2)= \frac{i}{2}\int dx (\varphi_1^* \pi_2-\varphi_2 \pi_1^*) $$
where
$\pi$ is the conjugate momentum from which the norm follows as $(\varphi, \varphi)$.
In order to evaluate whether the norm is positive or negative we insert $\varphi(x,t)=\Theta(x) e^{i k x} e^{-i \omega t}$ in the expression for the norm and obtain
$$ (\varphi, \varphi)= \frac{M_s^2}{\lambda |\gamma|^2}\int dx |\Theta(x)|^2  (\omega-|\gamma| H/M_s)$$
$$=\frac{M_s^2}{\lambda \gamma^2} \int dx |\Theta(x)|^2 \left( \pm \sqrt{ A \frac{ \lambda |\gamma|^2}{M_s^2}k^2+  K_z \frac{\lambda |\gamma|^2}{M_s^2}}  \right).$$
We therefore confirm the positive or negative sign of the norm. That is, the frequency $\omega_{++}$ has positive norm and the frequency $\omega_{+-}$ has negative norm, in agreement with what was found previously.

\section{Reflection and transmission coefficients}
In this section we explain how the reflection and transmission coefficients are obtained, following ~\cite{giacomelli2021superradiant,manogue1988klein}. 
We start writing the time-independent equation, by inserting $\varphi(x,t)=\varphi(x)e^{-i \omega t}$ into Eq.~(\ref{eqn: KleinAFM}). This results in
\begin{equation}
-  \frac{K_z |\gamma|}{M_s} \varphi(x) + \frac{A  |\gamma|}{M_s} \frac{d ^2\varphi(x)}{d x^2}+ \frac{M_s}{|\gamma| \lambda}(\omega- \gamma H/M_s)^2 \varphi(x)=0.
\end{equation}
We expand the solution of this equation in plane waves for an incoming wave from the left, i.e. 
\begin{equation}
\varphi(x)=
\begin{cases}
    = e^{i k_{-\infty}x}+ \mathcal{R} e^{-i k_{-\infty}x}\qquad \text{for}\quad  x \rightarrow -\infty ,\\
    =\mathcal{T} e^{i k_{+\infty}x}\qquad \text{for} \quad x \rightarrow + \infty\end{cases}
\label{asymp}
\end{equation}
with $\mathcal{R}$ and $\mathcal{T}$ reflection and transmission coefficients and where, explicitly, 
$A k_{+ \infty}^2=-K_z+(\bar{H}- \omega M_s / |\gamma|)^2/\lambda $
and $A k_{- \infty}^2=-K_z+( \omega M_s / |\gamma|)^2 / \lambda $.
This scattering ansatz is valid within the decay length of the spin waves, that scales as $1/\alpha$. Furthermore we choose $\varphi_1=\varphi$ and $\varphi_2=\varphi^*$. The detailed shape of the magnetic-field barrier affects the particular expression of the reflection and transmission coefficients, but not the relations among them, as we see in the following. In fact, if $\varphi_1$ and $\varphi_2$ are two general solutions of Eq.~(B1), then 
\begin{equation}
\varphi_1 (\partial_x \varphi_2)-(\partial_x \varphi_1) \varphi_2
\end{equation}
is a conserved quantity along $\hat{x}$, because the first derivative of this expression vanishes thanks to Eq.~(B1). Therefore we can equate this expression evaluated in any two positions. We choose two positions on opposite sides of the barrier (as in Eq.~$(\ref{asymp})$). Furthermore we choose $\varphi_1=\varphi$ and $\varphi_2=\varphi^*$. This yields 
$$ |\mathcal{R}|^2=1-\frac{k_{+\infty}}{k_{-\infty}}|\mathcal{T}|^2  .$$
From this formula, we notice that if $k_{+\infty}/k_{-\infty}>0$, $|\mathcal{R}|^2<1$, i.e. conventional scattering, while if $k_{+\infty}/k_{-\infty}<0$, then $|\mathcal{R}|^2>1$, meaning superradiant scattering. To determine the sign of the momentum $k_{\pm \infty}$, it is necessary to compute the sign of the group velocity, i.e we require $\partial_{k_{+\infty}}\omega \propto (k_{+\infty}/(\omega-\gamma \bar{H}))>0$, $\partial_{k_{-\infty}}\omega \propto (k_{-\infty}/\omega)>0$. 
Dividing the frequencies in ranges according to the behaviour of waves within those frequencies, with extremes $]-H_{\rm sf}, H_{\rm sf}, \gamma \bar{H}-H_{\rm sf}, \gamma \bar{H}+H_{\rm sf}[$,
we find $k_{+\infty}<0$ and $k_{-\infty}>0$ for the frequencies in the range $H_{\rm sf}<\omega M_s /|\gamma|<\bar{H}-H_{\rm sf}$. 
\section{ Large velocities}
Here, we can rederive the dispersion relation and sign of the norm, using the formalism typically used in analogue gravity~\cite{barcelo2011analogue,corley1999black,jacquet2020next}. In fact, usually in analogue gravity an effective metric is built in which the main elements are the background velocity and the sound velocity.
This approach gives us the advantage of including the polarized charge current in the calculation of the norm and to confirm our results in terms of positive or negative energy contributions.
Explicitly taking into account the background velocity, we can start from the Lagrangian describing the AFM
$$\mathcal{L}=\frac{M_s^2}{\lambda |\gamma|^2}|(\partial_t + v \partial_x)\varphi|^2    -A(\partial_x \varphi)(\partial_x \varphi^*)+K_z \varphi \varphi^*$$from which we compute the corresponding Euler Lagrangian equations of motion
$$ -K_z \varphi +A (\partial_x^2 \varphi)-\frac{M_s^2}{\lambda |\gamma|^2}(\partial_t+v \partial_x )(\partial_t+v \partial_x ) \varphi=0,$$We look for the scalar product~\cite{robertson2012theory,birrell1984quantum} for this Lagrangian 
$$(\varphi_1,\varphi_2)=\frac{i}{2}\int dx (\varphi_1^* \pi_2-\varphi_2 \pi_1^*),$$where $\pi$ is the conjugate momentum $\pi=\frac{\partial \mathcal{L}}{\partial \dot{\varphi}^*}=\frac{M_s^2}{|\gamma|^2 \lambda}(\partial_t+v \partial_x )\varphi$, which brings us to the norm (inserting spin wave solution $\varphi(x,t)= \Theta e^{i k x- i \omega t} $)
$$(\varphi,\varphi)= \frac{M_s^2}{|\gamma|^2 \lambda} \int dx |\Theta|^2 (\omega-  k_x v).$$
Inserting the dispersion relation results in
$$ (\varphi,\varphi)=\frac{M_s}{|\gamma| \lambda} \int dx |\Theta|^2 ( \pm \sqrt{A \lambda}|k|).$$
Therefore, even taking into account the polarized charged current explicitly, we obtain that $\omega_{++}$ has positive norm, in agreement with our calculations above.
Click Tidy to clean up the entries below      

\end{document}